\documentclass[prd,twocolumn,superscriptaddress,nofootinbib,amsmath,amssymb,aps,10pt]{revtex4-1}

\usepackage{soul}
\usepackage{graphicx}
\usepackage{dcolumn}
\usepackage{bm}
\usepackage[utf8]{inputenc}
\usepackage{amsmath, amsthm, amsfonts, amssymb}
\usepackage[svgnames]{xcolor}
\definecolor{DarkRed}{RGB}{179, 27, 27}
\colorlet{color1}{NavyBlue}
\usepackage[colorlinks=true,allcolors=DarkRed]{hyperref}
\usepackage{bm}
\usepackage{physics}
\usepackage{dsfont}
\usepackage{graphicx}
\usepackage{cleveref}
\usepackage{xfrac}
\usepackage{mathrsfs}
\usepackage{comment}

\def\apj{{ApJ}}

\def\apjl{{ApJL}}

\def\aap{{A\&A}}

\def\mnras{{MNRAS}}

\def\prd{{Physical Review D}}

\def\04a{{2004 a}}
\def\04b{{2004 b}}
\newcommand{\bx}{\boldsymbol{x}}

\begin{document}

\title{Random walks around black holes and low-frequency X-ray variability}
\date{\today}
\author{Arthur G. Suvorov}\thanks{a.suvorov@uni-tuebingen.de}
\affiliation{Theoretical Astrophysics, Eberhard Karls University of T\"ubingen, T\"ubingen 72076, Germany}

\begin{abstract}
\noindent The stochastic dynamics of a grain embedded within a turbulent fluid subject to strong gravitational fields can be formulated as a random walk on a Riemannian manifold.
Such curvature-weighted walks provide a framework to model the intrinsic variability of accretion onto compact objects.
By solving the relevant Fokker-Planck equation on a black hole background, we find the counterintuitive result that the escape probability of a grain is actually higher compared to flat space.
This is a consequence of the stretching of radial cells near the event horizon: there is a greater spatial volume for the particle to wander through before being captured.
By simulating a large number of grain trajectories, initially distributed on concentric shells with a density profile set by the thin-disc structure equations, we also study particle fluxes through the horizon.
Shallower spectral indices emerge at low frequencies relative to flat space, primarily due to time dilation, and steeper ones at high frequencies.
We find that Schwarzschild-weighted spectra broadly match observations of low-frequency X-ray variability from systems like Cygnus X-1 in their hard state, suggesting that geometric drifts may be important in describing stochastic accretion processes.
\end{abstract}

\maketitle


\section{Introduction}

Point particles in general relativity (GR) trace geodesics. 
On mesoscopic scales, however, grains embedded within fluid parcels may be subject to turbulent perturbations which cause them to move erratically \cite{pet01}.
Stochastic dynamics in this context are typically modelled via random walks or Brownian motion. 
When taking place on a curved space, there are innate geometric drifts that adjust capture probabilities and waiting time distributions \cite{grig98,hsu,ken87}.
Such adjustments may be important to account for when describing gas dynamics in astrophysical discs, for instance, to better understand X-ray variability in black-hole (BH) binaries \cite{kelly,wang09,mac10}.

Ensuring causality presents an immediate problem if one attempts to model relativistic, random diffusions.
In the standard setup of spatial random walks, the walker essentially flips a (weighted) coin at each time step to determine which direction they move in.
In a na{\"i}ve four-dimensional model, allowing diffusions in time means that the walker could theoretically fluctuate backwards in coordinate time, $t$, even while proper time, $\tau$, increases (i.e., their worldline may cease to be timelike).
To get around this problem, we consider a walker on a stationary background and ignore any self-gravity or backreaction (reasonable for a tiny dust grain surrounding a massive BH).
This allows us to set up the problem on a Riemannian hyperslice and to consider standard, spatial walks.

While the stochastic nature of turbulent grain dynamics have been considered in detail as regards X-ray variability in astrophysical systems (see, e.g., Refs.~\cite{van89,vaug03,yu22,chen23,fagin24} for a sample), the direct influence of spacetime curvature has not (to our knowledge).
The relevant equations are formulated in this paper, and studied at both the stochastic and Fokker-Planck levels. 
Importantly, thanks to time dilation, particles appear to slow down as they approach the horizon according to distant observers.
This stretches their arrival times and creates a pile-up of power at low frequencies, flattening the variability index (relative to a value $\sim 2$ for a flat-space walk) in a way that we argue better matches observations \cite{reig02}.
We demonstrate this explicitly by computing a large number ($\gtrsim 10^{7}$) of stochastic realisations in either flat space or over a Schwarzschild-de Sitter (SdS) background, keeping track of the number of walkers that fall beneath the event horizon to produce time series for the ``accretion rate''. {We carry this out either with a simple first-passage experiment or by repeatedly injecting particles into the outer edge of a fiducial disc to emulate steady-state accretion.}

The relevant equations describing random walks are derived in this short paper in the case of (slices of) static, spherically-symmetric spacetimes (Sec.~\ref{sec:formulation}). 
Worked examples are presented for escape probabilities (Sec.~\ref{sec:escape}), expected capture times (Sec.~\ref{sec:time}), and stochastic realisations (Sec.~\ref{sec:stochastic}).
Applications to X-ray variability from BHs are explored in Sec.~\ref{sec:accretion}, highlighting how curvature influences grain dynamics and modifies the spectral profile.
Some discussion is offered in Sec.~\ref{sec:conclusions}.

\section{Brownian motion in general relativity} \label{sec:formulation}

The concept of a random walk on a Riemannian manifold is well developed in the literature (see, e.g., Refs.~\cite{grig98,hsu}). 
We review the basic concepts here to pave the way for astrophysical investigations concerning walks around BHs in subsequent sections.

To define Brownian motion on a curved space, the following idea can be applied. 
When a grain receives a random kick in direction $a$ at time $t$, $d W^{a}_{t}$, it can thought of as selecting a direction on the tangent space at its current position.
Since the tangent space at any specific point on the manifold is flat, this allows us to ``lift'' Brownian motions to a curved framework (effectively mapping $\partial \to \nabla$).
Essentially, we suppose the walker receives a \emph{flat} random kick, and then use a frame transformation to translate that kick into a coordinate displacement on the curved position space \cite{grig98}.
Notably, Weiner kicks satisfy $dW^a_t dW^b_t = \delta^{ab} d\tau$, with $\tau$ being some affine parameter (i.e., proper time for a timelike grain) \cite{hsu}.

To carry out the above, we define ${W}_{t}$ as a standard Brownian motion in $\mathbb{R}^3$ (extensions to higher dimensions are straightforward). 
By introducing frame fields, $e^\mu_a(\bx)$, defined as
\begin{equation}
    g^{\mu\nu}(\bx) = \delta^{ab} e^\mu_a(\bx) e^\nu_b(\bx),
\end{equation}
we can write the stochastic differential equation describing the position of the walker in space\footnote{Note the factor 1/2 here is, in some sense, manually implemented to avoid the so-called ``It\^{o}-Stratonovich Dilemma'' to recover the expected diffusion limit; see footnote 2 in Ref.~\cite{grig98} and Ref.~\cite{man12} for a general discussion. In most of what follows we consider a constant diffusivity, however, and the ambiguity is moot.} \cite{ken87},
\begin{equation} \label{eq:strat}
\begin{aligned}
    dX^\mu_t &= \sigma(X_t) e^\mu_a(X_t) \circ dW^a_t \\
    &=\sigma e^\mu_a  dW^a_t + \frac{1}{2}\left[{\nabla^{\mu} \left(\sigma^2 \right)} + {\sigma^2} e^{\nu}_a \partial_{\nu} e^{\mu}_a  \right] d \tau \\
    &=\sigma e^\mu_a  dW^a_t + \frac{1}{2}\left[{\nabla^{\mu} \left(\sigma^2\right)} - {\sigma^2} g^{\alpha \beta} \Gamma^{\mu}_{\alpha \beta}  \right] d \tau.
    \end{aligned}
\end{equation}
{In this system of equations, $X_t$ is the state variable representing the position of the random walker at (coordinate) time $t$,} $g^{\mu \nu}$ is the metric tensor, the $\Gamma^{\mu}_{\alpha \beta}$ are the Christoffel symbols, $\circ$ denotes the Stratonovich operation, $\sigma$ is a diffusion coefficient (which can vary as a function of time and position in general), and Greek indices are associated with the manifold coordinates while Latin ones correspond to the local frame.
In the second line of equation \eqref{eq:strat}, we have made use of the It{\^o} conversion formula, $X \circ dY = X dY + \frac{1}{2} dX dY$.
In flat space, equation \eqref{eq:strat} reduces to the standard one describing Wiener processes.
Indeed, the first term on the right-hand side represents white noise, the second a diffusive drift, and the final a \emph{geometric drift} induced by curvature. 

\subsection{Large $N$ limit} \label{sec:largen}

In returning to our original problem of considering grains embedded within gas parcels on a BH spacetime, it may be impractical to track the motions of $N \gg 1$ grains as this would require solving equation \eqref{eq:strat} $N$ times.
While we \emph{do} solve the stochastic equations numerically in Secs.~\ref{sec:stochastic} and \ref{sec:accretion} in the SdS case, it is worth also considering the statistical limit where one models instead a probability density function, $P$, associated with large $N$ to gain intuition. 
This maps the inherent stochasticity into something deterministic via the familiar Fokker-Planck construction.
For any given (well-behaved) test function of the random position, $F(X_t)$, It{\^o}'s lemma gives $dF = (\partial_\mu F) dX^{\mu}_t + \tfrac{1}{2}(\partial_{\mu} \partial_{\nu} F) dX^{\mu}_t dX^{\nu}_t$. 
Taking the expected value of the differential $dF$, $\mathbb{E}[dF]$, allows us to define the generator, $\mathcal{L}$, from which the Fokker-Planck equation reads $0 = \partial_t P - \mathcal{L}^{\dagger} P$ where $\dagger$ denotes the Hermitian adjoint \cite{hsu}.
The expectation eliminates the noise term as $\mathbb{E}[dW_t] = 0$ by definition. The geometric drift, however, involves the contraction of two Wiener kicks; combining the above expansion with equation \eqref{eq:strat} yields
\begin{align}
    \mathbb{E}[dF] &= \frac{1}{2}\left[\left( \nabla^\mu \sigma^2 \right) \nabla_\mu F +{\sigma^2} \nabla_{\mu} \nabla^{\mu} F  \right] d \tau \\
    &= \frac{1}{2}\nabla_\mu \left( \sigma^2 \nabla^\mu F \right) d\tau.
\end{align}
This means that the generator is essentially a weighted Laplace-Beltrami operator,
\begin{equation} \label{eq:generator}
    \mathcal{L} = \frac{1}{2}  \nabla_\mu \left(\sigma^2 \nabla^\mu \right),
\end{equation}
as physically expected.

Since the Laplacian is self-adjoint\footnote{This property ensures that (i) that the forward and backward evolution operators are equal (i.e., no probability currents), and (ii) the solution, in the steady-state limit, converges to the uniform distribution.} with respect to the Riemannian measure $dV$, the Fokker-Planck equation reduces to a simple generalisation of the heat equation, viz.
\begin{equation} \label{eq:diffusioneqn}
0 =    \frac{\partial P}{\partial t} - \frac{1}{2} g^{\mu \nu} \nabla_{\mu} \left(\sigma^2 \nabla_{\nu}  P \right) .
\end{equation}
In the large deviation limit -- corresponding to a diffusionless gas where grains are free ($\sigma \to 0$) -- we recover the geodesic equation from \eqref{eq:diffusioneqn} (see Appendix~\ref{sec:geodesics}).

Note that in the context of the \citet{ss73} model we can estimate ${\sigma^2}/{2} \approx \alpha c_s H/{\rm Sc}$ where $\rm{Sc}$ is the Schmidt number, $c_s$ is the speed of sound, $H$ is the disc thickness, and $\alpha$ is the dimensionless viscosity coefficient.
In the astrophysical cases of interest, the Schmidt number is likely of order unity \cite{youd07} so that $\sigma$ could be estimated directly from the disc structure equations.

\section{Stumbling around a black hole} \label{sec:bhwalks}

At either the stochastic \eqref{eq:strat} or bulk \eqref{eq:diffusioneqn} levels, we return to our original problem of studying grain dynamics around BHs.
We consider a static, spherically-symmetric line element,
\begin{equation}
    ds^2 = -f(r)dt^2 + f(r)^{-1}dr^2 + r^2 d \Omega^2,
\end{equation}
in the usual Schwarzschild-like coordinates.
As noted earlier, we restrict attention to spacelike slices to avoid the diffusion into yesterday problem {(see also Appendix~\ref{sec:geodesics})}.
On the spacelike slice where our walk takes place, $\Sigma$, the relevant line element is $d\Sigma^2 = f(r)^{-1} dr^2 + r^2 d\Omega^2$. 

We focus here on the SdS case for concreteness, where we set $f(r) = 1 - \frac{2M}{r} - {\Lambda}r^2$ in geometric units. 
Provided that $\Lambda > 0$ and $27 \Lambda M^2 < 1$, the function $f$ possesses two positive roots bounding the spatial patch between the event horizon, $r_H$, and the outer cosmological horizon, $r_c$.
These represent two \emph{absorbing} boundaries: the walk terminates when the grain reaches either $r_H$ or $r_c$.
In the derivations below however, we keep the function $f$ general where possible so that the reader could instead consider their favourite compact object.
Indeed, if we reside within a universe with positive cosmological constant \cite{planck13}, the cosmological horizon ought to serve as an absolute outer boundary for our walker (in an astrophysical context, we could instead consider the outer edge of a disc).
Even in the limit of weak curvature, this changes the nature of the walk as absorbing boundary conditions apply \emph{at both ends}.
Notably, in unbounded Euclidean space, it is known that walks in $\geq 3$ dimensions are transient.
Such a theorem is often discussed in the humorous context of a ``stumbling drunk'': only in $<3$ dimensions can the walker find their way home with certainty, otherwise they escape every bounded region almost surely \cite{rick}.
In any case, the presence of an additional, absorbing boundary arguably increases the astrophysical realism of the problem. 

In the flat 3D case, because the spatial volume element grows like $r^2$, the walker is strongly biased to move outward. 
This encapsulates the ``drunk walker'' theorem: they stumble to infinity with probability 1 (i.e., the walk is transient).
In the present context, however, the hyperslice is bounded ($r_{H} \leq r \leq r_{c}$) and because there is no infinity to wander towards, the random walk is no longer transient but rather recurrent with respect to the boundaries.
In other words, the walker must eventually hit either $r_H$ or $r_c$, the respective probabilities of which we can estimate from the diffusion equation \eqref{eq:diffusioneqn}.

\subsection{Escape and capture probabilities} \label{sec:escape}

The probability $P_{H}(r)$ that the walker is swallowed by the event horizon before it reaches the cosmological horizon is a harmonic function satisfying $\mathcal{L}^{\dagger} P_{H} = 0$ (i.e., it is an equilibrium solution to equation \ref{eq:diffusioneqn} because the ultimate outcome of the journey is necessarily time independent) \cite{grig98}.
For a constant $\sigma$, we can immediately integrate the Laplacian once to find
\begin{equation}
\frac{\partial P_{H}} {\partial r} = \frac{c_1}{r^2 \sqrt{f(r)}},
\end{equation}
for some constant $c_1$.
Integrating again from $r$ up to $r_c$ and applying the absorbing boundary conditions, $P_{H}(r_H)=1$ and $P_{H}(r_c)=0$, yields the solution
\begin{equation} \label{eq:pHf}
    P_{H}(r) = \frac{\int_{r}^{r_c} du \, u^{-2} f(u)^{-1/2}}{\int_{r_H}^{r_c} du \, u^{-2} f(u)^{-1/2}}.
\end{equation}
Expression \eqref{eq:pHf} can be evaluated exactly using elliptic integrals in the SdS case, though the result is rather unwieldy from a numerical perspective.
In the small $\Lambda$ limit for SdS, some useful formulae obtained through Taylor expansion to leading order are
\begin{widetext}
\begin{equation} \label{eq:smalllambda}
\int \frac{du}{u^2 \sqrt{f(u)}} = -\frac{\sqrt{1 - \frac{2M}{u}}}{M} +\frac{\Lambda}{2} \left[ u \sqrt{1 - \frac{2M}{u}} + 3M \log\left(\sqrt{u} + \sqrt{u - 2M}\right) \right] + \mathcal{O}(\Lambda^2),
\end{equation}
\end{widetext}
$r_H \approx 2 M(1+ 4 M^2 \Lambda)$, and $r_c \approx 1/\sqrt{\Lambda} -M$. These can be applied reliably for $\Lambda \ll 1$.

\begin{figure}
\centering
  \includegraphics[width=0.485\textwidth]{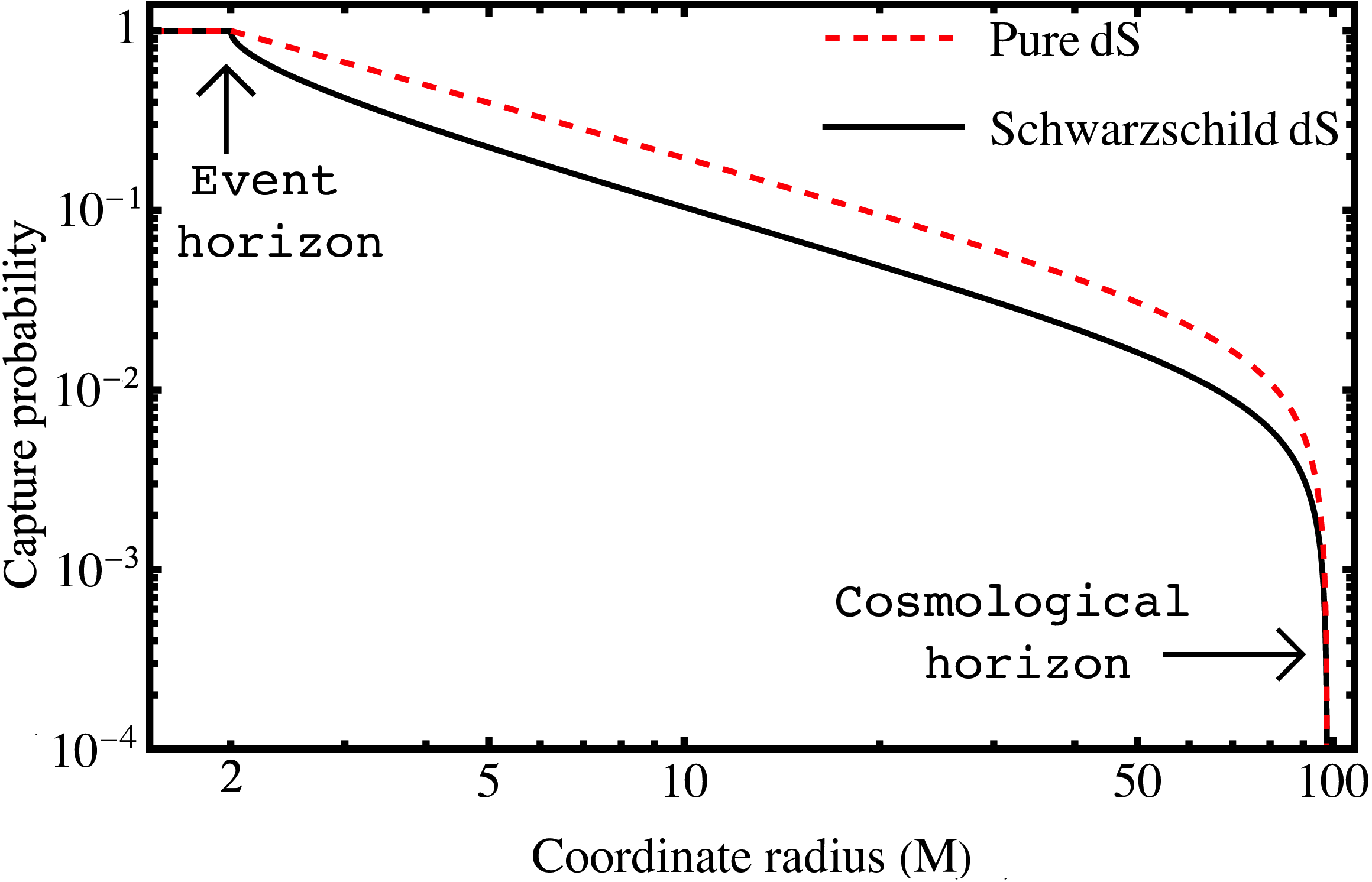}
\caption{Capture probability, $P_{H}(r)$ from equation \eqref{eq:pHf}, for a grain subject to diffusion on a Schwarzschild dS slice for $\Lambda = 10^{-4}$ and either {$M = 0$ (red) or $M = 1$ (black)}.
  }
  \label{fig:capture}
\end{figure}

Figure~\ref{fig:capture} depicts the function $P_{H}$ from \eqref{eq:pHf} for some (arbitrary) choices $\Lambda = 10^{-4}$ and either $M = 1$ (black) or $M=0$ (red; i.e., pure dS) in appropriate units.
(Even though there is no event horizon in the pure dS case, we can still place an absorbing boundary at the same $r_H$ to facilitate a direct comparison and demonstrate how a gravitational well influences the outcome).
We see that the probability of capture is monotonically decreasing as a function of radius, as expected, and approaches unity as $r \to r_H \approx 2 M(1 + 4 M^2 \Lambda)$.
Remarkably, in cases with non-zero mass the probability of capture is actually \emph{lower} for any given initial radius.
This occurs because the radial potential $1/f(r)$ weighting the integrals in expression \eqref{eq:pHf} grows sharply as $r \to r_H$, thereby stretching the proper distance as one approaches the BH. 
This is a physical consequence of redshift: there is a greater spatial volume near $r_H$ for the particle to wander through without collision compared to the dS or flat cases, and thus it escapes more often.
This represents a fundamental difference between the single particle case (i.e., geodesics) and turbulent grain dynamics.
For example, for a grain placed at $r \sim 5 M$ there is only a $\sim 20\%$ chance that it eventually wanders past the event horizon in this particular example: this lies beneath the innermost stable circular orbit (ISCO) where classical particle trajectories are inherently unstable.

\subsection{Timescales} \label{sec:time}

\noindent Having established a grain may succumb to two distinct fates through equation \eqref{eq:pHf}, we may ask how long it takes for either outcome to occur.
Thanks to the Feynman-Kac formula, we can compute the expected proper time, $\mathbb{E}[\tilde{\tau}]$, spent in the patch $r_H \leq r \leq r_c$ by solving $\mathcal{L}^{\dagger} \mathbb{E}(\tilde{\tau}) = -1$ (see Eq.~2.6 in Ref.~\cite{cab08}). That is, we must solve the inhomogeneous equation
\begin{equation} \label{eq:expectedeqn}
0 =    \frac{\partial}{\partial r} \left( r^2 \sqrt{f(r)} \frac{\partial \mathbb{E}[\tilde{\tau}]}{\partial r} \right) +\frac{2r^2}{\sigma^2 \sqrt{f(r)}},
\end{equation}
subject to the boundary conditions $\mathbb{E}[\tilde{\tau}(r_H)] = 0 = \mathbb{E}[\tilde{\tau}(r_c)]$ (i.e., if the grain already resides at an absorbing boundary it stays there).
Equation \eqref{eq:expectedeqn} admits a solution via quadrature,
\begin{equation} \label{eq:expectedsoln}
    \mathbb{E}[\tilde{\tau}(r)] = \int_{r_H}^r \frac{du}{u^2 \sqrt{f(u)}} \left( c_2 - \frac{2}{\sigma^2} \int_{r_H}^{u} dx \frac{ x^2}{\sqrt{f(x)}} \right),
\end{equation}
where the constant $c_2$ is given by
\begin{equation} \label{eq:c2}
    c_2 = \frac{2}{\sigma^2} \frac{\int_{r_H}^{r_c} \frac{dw}{w^2 \sqrt{f(w)}} \left( \int_{r_H}^w dx\frac{x^2}{\sqrt{f(x)}} \right)}{\int_{r_H}^{r_c} \frac{dw}{w^2 \sqrt{f(w)}}}.
\end{equation}
Unfortunately, the arguments resist integration in terms of known functions in general and a numerical treatment is necessary.
Alternatively, we can apply Taylor expansions, as in expression \eqref{eq:smalllambda}, to obtain something analytic (albeit cumbersome) in the small $\Lambda$ limit.

\begin{figure}
\centering
  \includegraphics[width=0.485\textwidth]{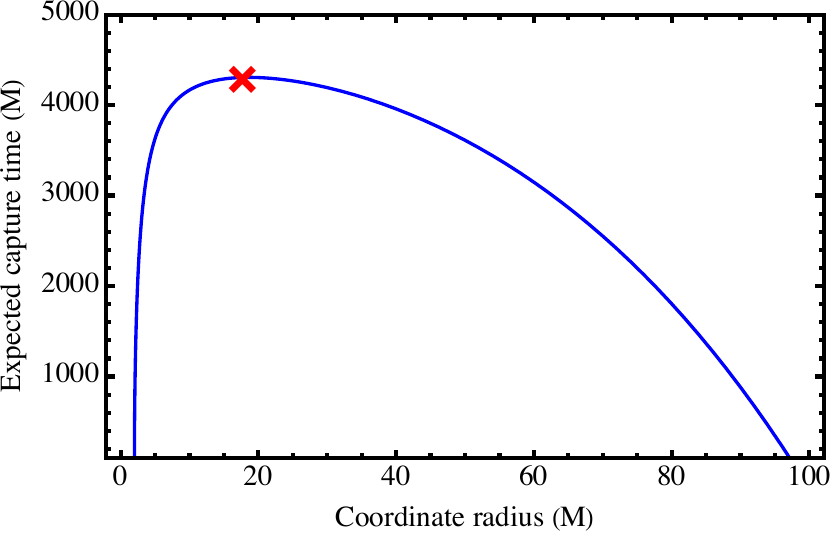}
  \caption{Expected capture time by either horizon, expression \eqref{eq:expectedsoln}, for $M = 1$, $\Lambda = 10^{-4}$, and $\sigma = 1$ (in appropriate units). The global maximum at $r_{\rm max} \approx 18.6 M$ is highlighted by a red cross.
  }
  \label{fig:time}
\end{figure}

Figure~\ref{fig:time} depicts the small-$\Lambda$ expected waiting time \eqref{eq:expectedsoln} for the choices $M = 1$, $\Lambda = 10^{-4}$, and $\sigma = 1$ (again in some appropriate units). 
When the particle is near either boundary the expectation plummets to zero, though generally displays a non-trivial structure.
For instance, the maximum occurs at $r_{\rm max} \approx 18.61 M$ with $\mathbb{E}[\tilde{\tau}(r_{\rm max})] \approx 4306$. 
This indicates that the peak of the competition set by the two absorbing boundaries does not, in general, take a simple value but depends on the geometric tugging from the two horizons.
Moreover, the capture time hosts especially strong gradients in the neighbourhood of $r_H$. 
This reflects the structure of the probability distribution \eqref{eq:pHf}: it becomes unlikely for the grain to be swallowed by the event horizon as it moves away (Fig.~\ref{fig:capture}) but, since $r_c \gg r_H$, the particle must wander for a long time before ending its journey.
The same qualitative behaviour is observed for any $M$ or $\Lambda \ll 1$, with $\mathbb{E}[\tilde{\tau}]$ scaling simply as $\sigma^{-2}$ as is clear from equations \eqref{eq:expectedsoln} and \eqref{eq:c2}.

\subsection{Stochastic realisations} \label{sec:stochastic}

Rather than considering diffusive and large-deviations limits, we can alternatively simulate random walks directly from equation \eqref{eq:strat}.
By introducing a discretised radial step, $\Delta r$, we apply standard numerical methods to visualise how a grain may jolt about for any given metric and parameter choices.
We use the simple Euler-Maruyama method with a fixed step $\Delta r = 0.1 M$, placing the grain initially at $r = 10 M$ in a setup with $M = 1$, $\Lambda = 10^{-4}$, and $\sigma = 1$ in appropriate units.

\begin{figure*}
\centering
  \includegraphics[width=0.9\textwidth]{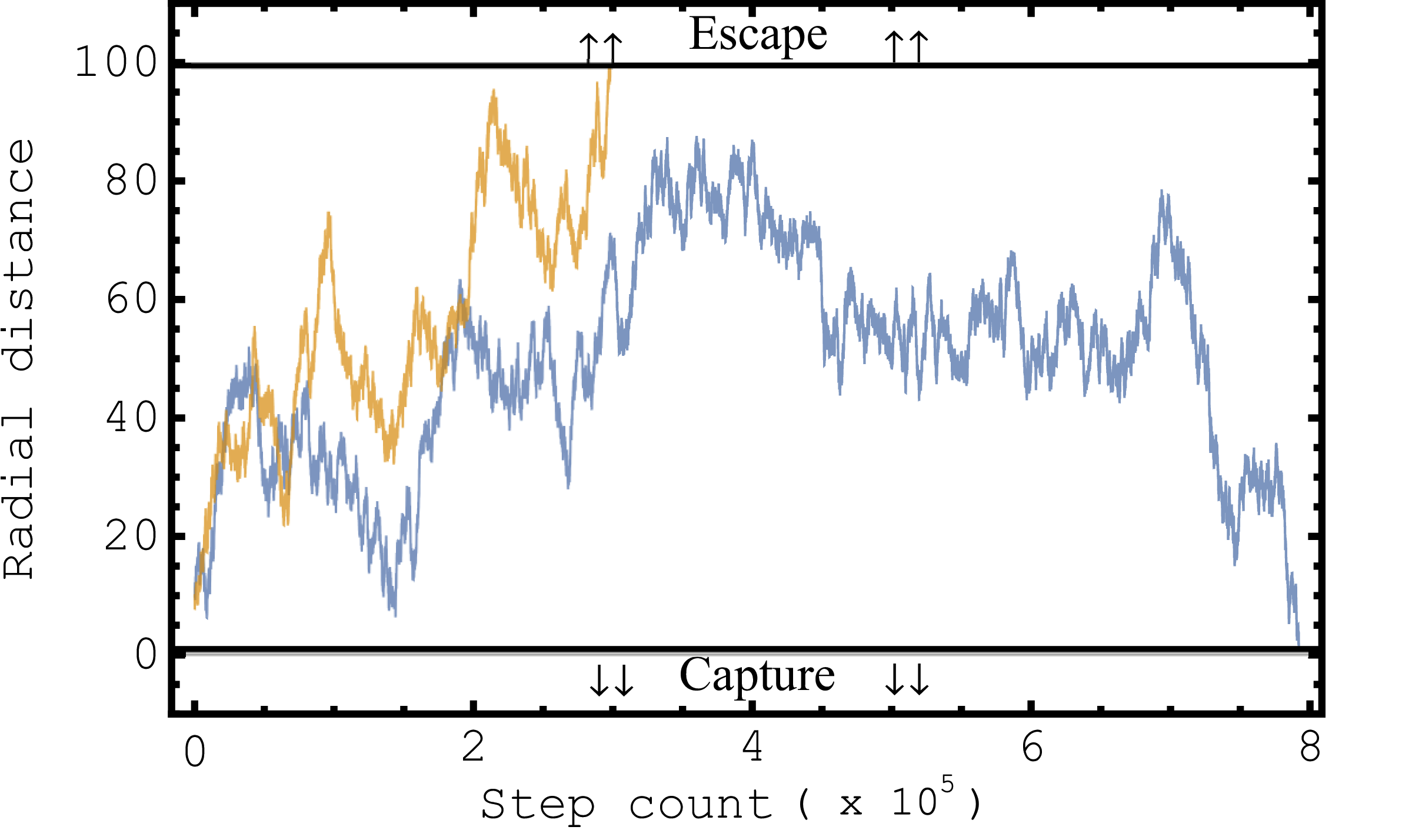}
  \caption{Two possible realisations of a grain beginning at a position of $r_{\rm init} = 10 M$, culminating in either escape through $r_c$ (orange) or capture through $r_H$ (blue), for random walks around an SdS BH with $M = 1$, $\Lambda = 10^{-4}$, and $\sigma = 1$ (Equation~\ref{eq:strat}). The numerical step size is fixed as $\Delta r = 0.1 M$.
  }
  \label{fig:escape}
\end{figure*}

Two realisations, showing either escape through $r_c$ (orange) or capture at $r_H$ (blue), are depicted in Figure~\ref{fig:escape}.
In the case of eventual capture, we see that the particle nearly wanders through $r_c$ after $\sim 4 \times 10^{5}$~steps but gets pulled back after $\sim 7 \times 10^{5}$~steps, ultimately being consumed by the event horizon. 
As can be seen from Fig.~\ref{fig:capture}, such outcomes are relatively rare (occurring in only $\sim 10\%$ of cases for these initial conditions). 
Since our random walk is subject to no probability currents (Footnote 2), the process is memoryless in general: the coin flips at any given step do not depend on time explicitly and the walker forgets where they were previously.
Figure~\ref{fig:escape} essentially represents the possible dynamical outcomes of a single grain embedded within a turbulent gas parcel in a universe consisting of a single SdS BH.

\section{Implications for X-ray variability} \label{sec:accretion}

Based on the stochastic formulation, we can model $N \gg 1$ grains to make a prediction for the accretion rate by attaching a mass to each grain and counting the number falling through $r_H$.
To achieve this, we note that the stochastic process on the Riemannian slice satisfies (It{\^o}'s lemma)
\begin{equation}
    d X_t^\nu dX_t^\mu = \sigma^2 g^{\mu \nu} d \tau,
\end{equation}
which implies that (from the $rr$-component)
\begin{equation} \label{eq:geoweight1}
    dr^2 = {\sigma^2 f(r)} {d\tau}.
\end{equation}
A simple rearrangement gives
\begin{equation} \label{eq:geoweight}
        d\tau = \frac{dr^2}{\sigma^2 f(r)} ,
\end{equation}
and thus we can attach a physical duration to any given walk that terminates at $r_H$ or otherwise.
For radial walks, the coordinate time seen by a distant observer satisfies
\begin{equation} \label{eq:dtau}
    d \tau = \sqrt{f(r)} dt,
\end{equation}
allowing us to keep track of the observer-identified particle flux as a function of time.
We track the geometric weight implied by expressions \eqref{eq:geoweight} and \eqref{eq:dtau} for each individual particle that reaches close to the horizon (i.e., the absorbing boundary, numerically set as $r_{\rm in} = r_H + \Delta r$; see Appendix~\ref{sec:convergence}) to get a running total of the time it takes to pass through the inner edge of the simulation domain.

\begin{figure*}
\centering
  \includegraphics[width=0.485\textwidth]{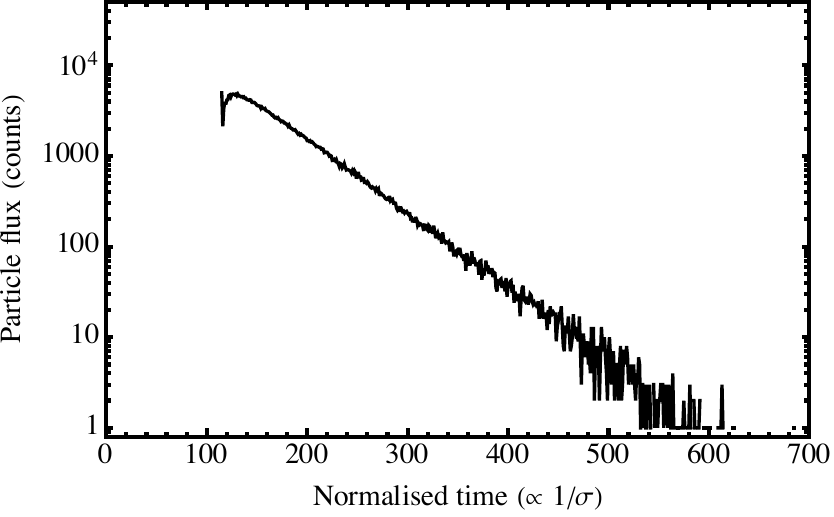}
    \includegraphics[width=0.485\textwidth]{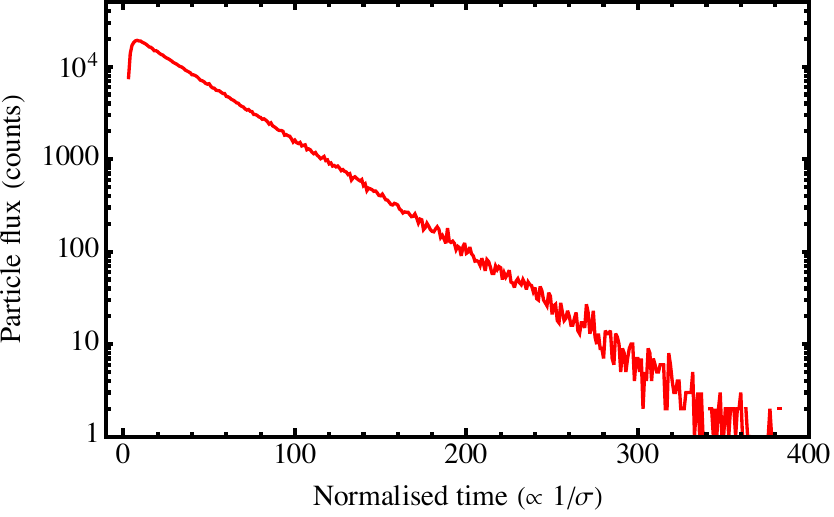}
\caption{Particle fluxes through the inner boundary as a function of normalised time for either the SdS (left, black) or flat [right, red; i.e., $f(r) = 1$] cases for $N = 3 \times 10^7$ grains beginning walks on various radial shells with a number density following equation \eqref{eq:particledensity}. Note the different horizontal, but equal vertical, scales.
  }
  \label{fig:accrates}
\end{figure*}

For a Keplerian disk, the number density of particles, of mass $m$, in the \citet{ss73} model scales as
\begin{equation} \label{eq:particledensity}
    n(r) \approx \frac{\langle\dot{M} \rangle}{3 \pi \alpha m} \left(\frac{H}{r}\right)^{-3} r^{-3/2},
\end{equation}
where $\langle \dot{M} \rangle$ is some average accretion rate. If we assume $H/r$ is constant -- as is standard in thin-disc modelling (e.g., \cite{gs21}) -- we obtain simply $n(r) \propto r^{-3/2}$.

{In what follows, it is worth noting that SdS geometries generally possess both an ISCO and an outermost-stable circular orbit (OSCO).
In our convention, these are respectively given by the smallest and largest real roots of the quartic \cite{stuk1,stuk2}
\begin{equation} \label{eq:iscoosco}
0= 4 \Lambda r^4 - 15M \Lambda r^3 - M r + 6 M^2,
\end{equation}
associated with the turning points of the effective potential computed from the spacetime metric [i.e., the roots of $0 = V''_{\rm eff}(r)$].
This contrasts with the case shown in Fig.~\ref{fig:escape} where the outer boundary was set as the cosmological horizon.
In the case of accretion discs, using the OSCO as the outer boundary is more physical (though, for astrophysically-realistic parameters, discs will likely be truncated at radii $r \ll r_{\rm OSCO}$). 
}

\subsection{First-passage fluxes} \label{sec:firstpassage}

{We begin by first examining the simple case of a first-passage experiment, initiating walks on a nested set of radial shells and counting those that fall beneath the horizon.
A case more appropriate for steady-state accretion is considered instead in Sec.~\ref{sec:steadystate}, where we instead inject particles at the outer edge of the disc.}

We use the particle density \eqref{eq:particledensity} to build $N_i(r_i)$ profiles for the number of walks beginning on shells of initial radius $r_i$ for a given total number of walkers $N$ [i.e., $\sum_i N_i(r_i) = N$].
More precisely, we expect a distribution $d N/dr \sim 2  \pi r H(r) n(r)$ for a thin disc and thus use $d N/dr \propto r^{1/2}$ to set the raw counts on each shell.
Particles are dropped at radial shells in spacings of $\Delta r$ starting from the ISCO {up to the OSCO.}
A flag records whether they escape (reach $r_{\rm out}$) or contribute to the accretion (reach $r_{\rm in}$) and the walk terminates.
The cumulative times of each particle from equation \eqref{eq:geoweight} are tracked to produce a time series for the horizon flux from the set of walks.
We set $N = 3 \times 10^{7}$ for the total number of particles to be distributed within the radial shells spaced equally by $\Delta r = 0.5 M$.
We have verified that adjusting the outer/inner boundaries or $\Delta r$ slightly does not qualitatively impact the results {and that some expected, theoretical scaling relations are recovered as we adjust them (see Appendix~\ref{sec:convergence})}.

The resulting histograms for {horizon-hitting particle fluxes} are depicted in Figure~\ref{fig:accrates}, where either the curvature drift is included (black; left) or not (red; right).
Two major effects are visible. 
First, the raw counts in the SdS case are lower than the flat one.
This is a consequence of the higher escape probability studied in Sec.~\ref{sec:escape} meaning that, for a given $N$, fewer particles enter the absorbing inner boundary and thus the overall flux is lower.
Secondly, for a fixed $\sigma$, the coordinate time it takes for particles to cross the inner boundary is longer in the SdS case (compare the horizontal axes).
{Though not shown, we have checked that increasing the value of $\Lambda$, and hence extending $r_{\rm OSCO}$, leads to more stretching as particles wander for longer and thus accumulate a greater net time-dilation.}
Time dilation also has the impact of steepening the slope of flux vs. time, primarily because the weighting \eqref{eq:geoweight} is most impactful in the final few steps, thereby reducing the overall variance of capture times.

\begin{figure*}
\centering
  \includegraphics[width=0.9\textwidth]{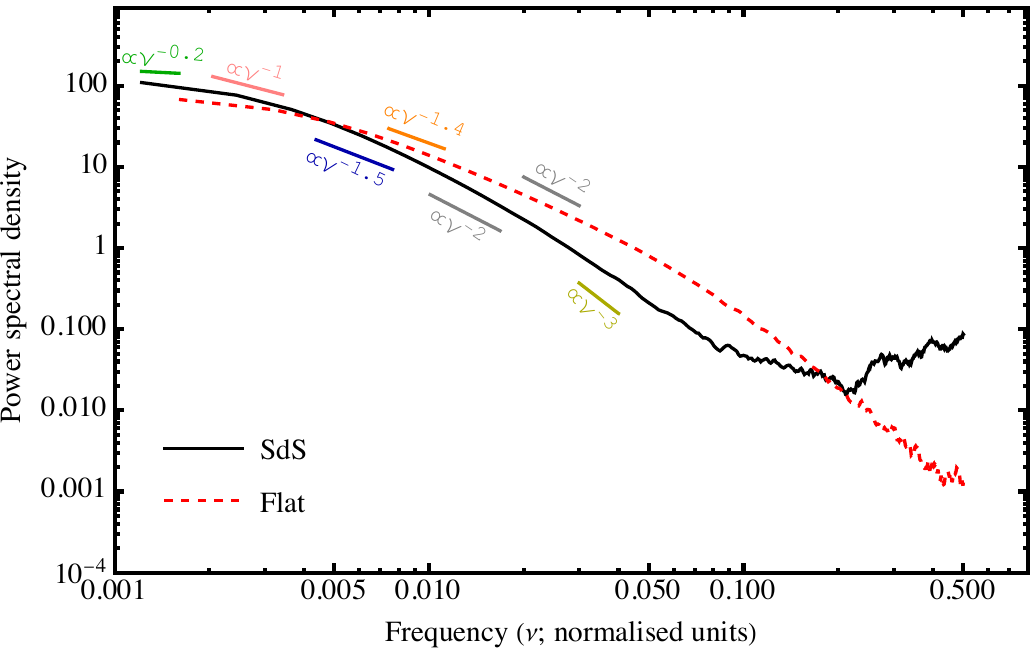}
\caption{Power spectrum as a function of normalised frequency ($\nu$) for the two ``accretion rates'' depicted in Fig.~\ref{fig:accrates}. Overlaid  are various power-law slopes to illustrate the steepening as a function of frequency.
  }
  \label{fig:psd}
\end{figure*}

Thanks to the geometric drift, the slope for the power spectral density (PSD) associated with flux variability changes because coordinate time scales like $f(r)^{-3/2}$ (equations \ref{eq:geoweight} and \ref{eq:dtau}), meaning that the fluctuations are essentially redshifted. 
This acts like a kind of non-linear low-pass filter relative to the flat space case.
This effect is demonstrated in Figure~\ref{fig:psd} showing the PSD for the two cases depicted in Fig.~\ref{fig:accrates}.

For the SdS case (black), we see a remarkably flat spectrum at very low frequencies (best-fit slope $\approx -0.2$).
This is simply because time dilation increases the accretion timescale (compare horizontal axes in Fig.~\ref{fig:accrates}), and thus the process develops power at lower frequencies (i.e., there is a longer period variability).
As the frequency increases the slope transitions to $\approx -1$ in both cases, reminiscent of flicker noise (see, e.g., Ref.~\cite{lyu97}).
In the ``intermediate'' regime the effect of curvature appears actually to \emph{steepen} the curve slightly; the flat-space index is $\approx -1.4$ while that for SdS is $\approx -1.5$ around $\nu \sim 0.003$ in the normalised units considered here.
The PSD continues to steepen, down to a slope of $\approx -3$ for SdS and $\approx -2$ for the flat case, until eventually showing uncorrelated noises, likely due to finite sampling effects.
{The finite-sampling limitation at high frequencies is most obvious in the SdS case because fewer particles enter the horizon as the capture probability is lower.}
{Either way,} the overall conclusion is that curvature naturally introduces a very shallow slope at low frequencies, which transitions to a steeper one at higher frequencies. 

This is in broad agreement with observations of flux variations in BH X-ray binaries.
A consistent trend in this respect is that, in the hard state, an almost-flat PSD is observed at the lowest frequencies, followed by a gradual steepening until reaching a noisy floor at the highest frequencies (see Ref.~\cite{van06} for a general overview).
For example, figure 2 in \citet{now00} showing fits for the source GX 339--4 depicts a power-law index of $\approx 0.1$ at frequencies $10^{-2} \lesssim \nu/\text{Hz} \lesssim 10^{-1}$ followed by a gradual steepening ($\gtrsim 2$ at $\nu \gtrsim 1$~Hz).
Similar trends are observed for the source GRO J1655--40 (see figure 5 in Ref.~\cite{remi99}, for instance).
Comparing to Fig.~\ref{fig:psd}, we see this qualitative behaviour matches better with the curvature-weighted walk relative to the flat-space one.
For the sources GRS 1915+105 and Cygnus X-1, \citet{reig02} found best-fit PSD indices of $\alpha = -1.12 \pm 0.04$ and $\alpha = -0.93 \pm 0.05$ at low-to-mid frequencies, respectively (see also Ref.~\cite{jiang24}).
The latter in particular, being shallower than $-1$, could be theoretically attributed to the geometric effects described above in the first transition region (at $\sim 10^{-3}$ in the normalised units considered here).

{We caution the reader however that we cannot directly compare to the observed X-ray variability of a source like Cygnus X-1, which is measured from light curves rather than particle counts.
In modern accretion models, some fraction of emissions originate away from the horizon (e.g., in the corona or jet base), especially in the hard state, and the observed variability can be amplified or filtered relative to the horizon flux depending on the ``efficiency'', $\eta,$ serving as the standard conversion factor between luminosity and accretion rate ($L_{\rm X}  \propto \eta\dot{M}$).
Moreover, In disc-fed accretors, the mass-capture process is more realistically represented through a continuous flow of particles entering the Lagrange point rather than through the diffusion of particles dropped simultaneously from a nested set of radial shells. 
We explore such a setup next.}

\subsection{Steady-state accretion} \label{sec:steadystate}

\begin{figure}
\centering
  \includegraphics[width=0.487\textwidth]{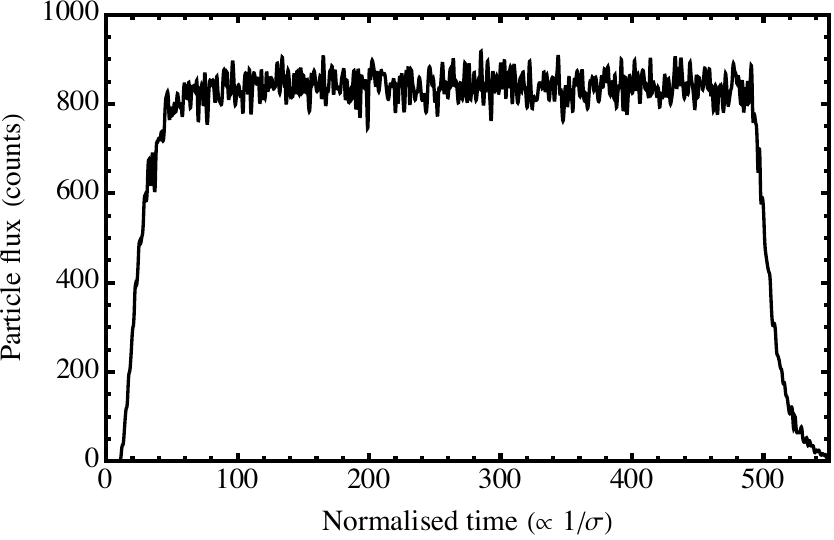}
\caption{Similar to Fig.~\ref{fig:accrates}, but instead for a ``steady-state'' process where particles are fed into the outer edge of the disc (i.e., just before the OSCO) in constant time intervals.
  }
  \label{fig:accsteady}
\end{figure}

{To better represent accretion in a (quasi-)steady state, we can instead populate the numerical shell before the OSCO (i.e., $r_i = r_{\rm OSCO}-\Delta r$) with $N_{i}$ walkers.
After a time $\Delta t$ we initiate another set of $N_{i}$ walkers from the same shell, serving as a proxy for mass inflow from the disc edge, repeating the process until the system reaches quasi-equilibrium with respect to inner-horizon fluxes.
Fixing $\Delta r = 0.5 M$ and $\Delta t = 2$ (in units such that $\sigma = 1$), the results of such a numerical experiment, using 2000 injections of $N_{i} = 10^{5}$ particles resulting in a global count of\footnote{A larger $N$ is necessary since the majority of walkers simply escape through the OSCO in accord with Fig.~\ref{fig:capture}.} $N = 2 \times 10^{8}$, are shown in Figure~\ref{fig:accsteady}.}

\begin{figure}
\centering
  \includegraphics[width=0.487\textwidth]{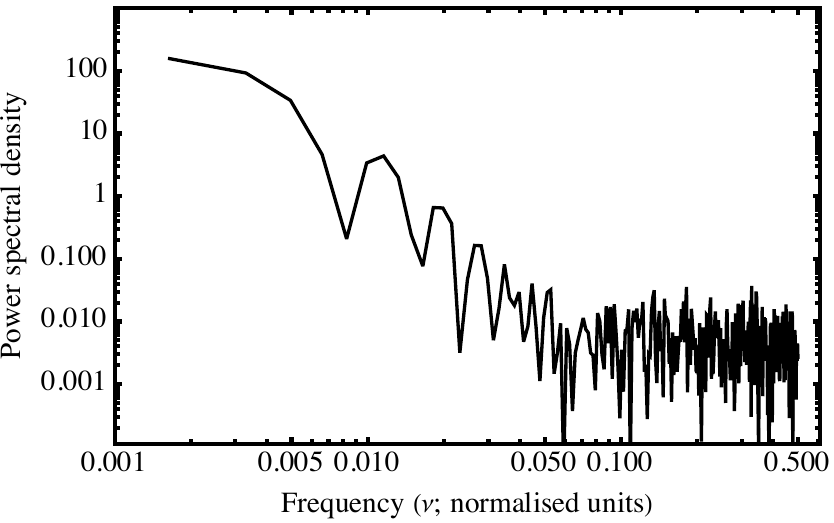}
\caption{Similar to Fig.~\ref{fig:psd}, but instead for the ``steady-state'' process depicted in Fig.~\ref{fig:accsteady}.
  }
  \label{fig:psdsteady}
\end{figure}

{After a rise time lasting $\sim 70$ units, we see that a baseline flux of $\approx 838$ particles per unit time is established (i.e., the mean).
Some time after the final injection is performed, the particle flux asymptotes to zero as the last stray walkers that have not escaped wander through $r_{\rm in}$.
Chopping the first and last $100$ units of the time series, we find a standard deviation of $\approx 29.5$ particles.
This corresponds to a consistent, fractional variability in the accretion rate of $\approx 4\%$ over the baseline.
Occasionally, a surplus of walkers fall through the horizon within the same time bin (e.g., around $\approx 300$ units of time in this particular realisation) leading to root-mean square variations of order $\sim 10\%$.
The typical variability is systematically lower in flat-space cases because walkers reach the horizon more easily (Fig.~\ref{fig:escape}) and tail events are washed out as meandering paths do not accrue any time dilation.
We show the PSD associated with this particular run in Figure~\ref{fig:psdsteady}.
Although correlated power jumps emerge at mid-to-high frequencies -- likely related to persistent injections of the same number of walkers with the same $\Delta t$ -- the same qualitative trends seen in Fig.~\ref{fig:psd} persist at the low end (i.e., a gradually steepening index as a function of frequency).}

{In its hard state, Cygnus X-1 shows typical variability on the order of $\sim 30\%$ \cite{axel18} which is a factor of several higher than the model.
However, many systems in the soft state -- such as GRO J1655--40 \cite{utt15} -- can show smaller variations of order a $\sim$~few percent, which is more consistent with our findings.
How the inclusion of additional physics, such as rotation or pre-horizon emissions discussed in the previous section, affects the picture discussed here is left to future work.
Nevertheless, while a thorough comparison lies beyond the scope of this paper, our findings suggest that curvature drifts may be important to account for in studies of the low-frequency X-ray variability of compact objects (provided the horizon flux provides a reasonable proxy for the accretion rate).}

\section{Discussion} \label{sec:conclusions}

In this paper, we have investigated how spatial random walks on curved manifolds can be used to study stochastic grain dynamics around BHs. 
For example, many such grains may exist in astrophysical discs around BHs, and their individual motions may set the variable accretion rate in the limit of strong turbulence.
Although random walks on curved spaces have been studied at length in the literature from a mathematical perspective, how the curvature drift impacts on the anticipated accretion rate has not, to our knowledge, been investigated before.
We have attempted to present material in a pedagogical way at either the Fokker-Planck (Secs.~\ref{sec:escape} and \ref{sec:time}) or stochastic (Sec.~\ref{sec:stochastic}) levels such that the relevant equations can be easily adapted to more realistic models in future (e.g., involving damped random walks \cite{yu22}).

It is important to stress that we have not attempted any serious astrophysical connections here. 
Our findings, however, uncover the counterintuitive result that the BH does not really act as a sink: the probability of capture actually decreases as $M$ increases (Fig.~\ref{fig:capture}) thanks to length contraction, leading to noisier capture (Figs.~\ref{fig:escape} and \ref{fig:accrates}).
The fact that the presence of a central mass decreases the probability of capture indicates that escape is generally preferable relative to the flat-space case (see also Fig.~\ref{fig:time}).
Moreover, by simulating a large number of walks and accounting for time dilation (Sec.~\ref{sec:accretion}), we found that the spectrum associated with particle flux through the horizon is generally shallower (steeper) at low (high) frequencies when compared to the flat case (see Fig.~\ref{fig:psd}). 
{Considering instead models of continuous injections (see Figs.~\ref{fig:accsteady} and \ref{fig:psdsteady}) leads to broadly similarly conclusions, though the typical variabilities are at most of order $\sim 10\%$.}
These results (though again based on a very simple model), appears to better match observations of low-frequency variability in X-ray binaries relative to flat-space walks \cite{reig02}.

Moving to more realistic cases requires, amongst other things, the use of a rotating solution. 
While more mathematically involved, diffusions over stationary geometries still admit the Laplace generator \eqref{eq:generator} \cite{grig98},
so that conceptual extensions are straightforward.
In the equatorial plane ($\theta = \pi/2$) of a Kerr BH, an additional advection term is present.
In conceptual terms, we have $\mathcal{L}_{\rm Kerr} \sim \Delta_{\rm SdS} + \omega(r) \partial_\phi$ where $\omega = -g_{t\phi}/g_{\phi\phi}$ is the angular velocity of the frame dragging. 
This dragging would cause the grain to circle around the BH and further bias the walk to longer timescales, much like water molecules draining in a bathtub.
It would be interesting to study such effects in more detail in future.
An even more bold attempt may be made to relate the stochastic dynamics to the spacetime multipole moments directly in an effort to consider a broad class of compact objects rather than on a case-by-case basis (though cf. Ref.~\cite{sg26}).
Including electromagnetic drifts would also be interesting in the context of neutron-star primaries (see, e.g., Ref.~\cite{scott97}).


\section*{Acknowledgements}
{I thank the anonymous referees for their feedback which prompted a much richer study than the original submission.}
I am grateful for support provided by the European Union's Horizon MSCA-2022 research and innovation programme `EinsteinWaves' 
under grant agreement No. 101131233, the Deutsche Forschungsgemeinschaft individual research grant 570901071, and the High Performance and Cloud Computing Group at the Zentrum f{\"u}r Datenverarbeitung of the University of T{\"u}bingen (Project `AnqaGW').

%

\appendix

\section{The point-particle limit} \label{sec:geodesics}

We rewrite the probability density, $P$, in terms of a new variable, $S$, that represents the mean of a Gaussian,
\begin{equation} \label{eq:somep}
    P(\bx, t) = \exp\left[ -\frac{S(\bx, t)}{\sigma^2} \right].
\end{equation}
This form is chosen so that intuitively, as $\sigma \to 0$, the probability ``cloud'' $P(\bx,t)$ obeying equation \eqref{eq:diffusioneqn} collapses into a delta function representing the localised position of a particle that ought to now follow geodesics. 
As we will see below, $S$ essentially represents the classical position and $\nabla_{\alpha} S$ plays the role of momentum. 

Considering a constant $\sigma$ and substituting the ansatz \eqref{eq:somep} into the diffusion equation \eqref{eq:diffusioneqn}, rearranging terms, and multiplying by $-\sigma^2$, we find
\begin{equation}
    \frac{\partial S}{\partial t} + \frac{1}{2} g^{\alpha \beta} (\nabla_\alpha S)(\nabla_\beta S) = \frac{\sigma^2}{2} \nabla^{\mu} \nabla_{\mu}S .
\end{equation}
Obviously, the right-hand side vanishes in the deterministic limit $\sigma \to 0$. We are thus left with a non-linear, first-order equation,
\begin{equation} \label{eq:classical}
0 =    \frac{\partial S}{\partial t} + \frac{1}{2} g^{\alpha \beta} (\nabla_\alpha S)(\nabla_\beta S) .
\end{equation}
Equation \eqref{eq:classical} is nothing but the Hamilton-Jacobi equation, $0 = \partial_t S + H(\bx, \boldsymbol{p})$ with $H(\bx, \boldsymbol{p}) = \frac{1}{2} g^{\mu \nu} p_\mu p_\nu$ for the classical momenta $p^\alpha \equiv d x^{\alpha} / d \tau = \partial^\alpha S$. 
By Hamilton's equations of motion, the trajectories that extremise the ``action'' $S$ are thus precisely the geodesics,
\begin{equation} \label{eq:3D_base_geo}
0=    \frac {d^2 x^\mu}{d \tau^2} + \Gamma^\mu_{\alpha \beta} \frac{ dx^\alpha}{d\tau} \frac{ d x^\beta}{d \tau}.
\end{equation}
As such, in the limit of zero noise, the probability density $P$ collapses in such a way that demonstrates, as hoped, free particles trace geodesics associated with the metric. 

{
The reader may notice however that we have only recovered the 3D geodesics in such a limit, rather than those associated with the spacetime.
At least in the case of static spacetimes (though see also Ref.~\cite{bies94} for stationary cases), however, it is possible actually to recover the full 4D geodesics using the ``Jacobi-metric'' method  \cite{ong75,gibbons16}.
We sketch such a reconstruction here, and its limitations from a stochastic perspective, for completeness.}

{As frame fields are not needed in this section, we switch to a more conventional notation where Latin indices represent spatial coordinates to avoid confusion with spacetime indices. 
Starting from some 3-metric $g_{ij}$, we introduce a conformal rescaling\footnote{ Brownian motion is strictly preserved under conformal transformations only in 2 dimensions \cite{levy40}.} through
\begin{equation} \label{eq:confmap}
\tilde{g}_{ij} = \Omega^2 g_{ij},
\end{equation}
for some sufficiently well-behaved function $\Omega$.
Using standard results for how the Christoffel symbols transform \cite{waldbook}, the geodesic equations for $\tilde{g}_{ij}$ read
\begin{widetext}
\begin{equation}\label{eq:3D_expanded}
0 = \frac{d^2 x^i}{d\lambda^2} + {}^{(g)}\Gamma^i_{jk} \frac{dx^j}{d\lambda} \frac{dx^k}{d\lambda} + 2 \frac{dx^i}{d\lambda} \frac{d(\ln \Omega)}{d\lambda} - \left( g_{j \ell} \frac{dx^j}{d\lambda} \frac{dx^\ell}{d\lambda} \right) g^{ik} \partial_k (\ln \Omega),
\end{equation}
\end{widetext}
where $\lambda$ is an affine parameter associated with $\tilde{g}_{ij}$; we identify $\lambda = \Omega^2 \tau$ to absorb non-affine acceleration terms in the rescaled geodesic equations noting that $\tau$ is an affine parameter associated with $g_{ij}$ \cite{waldbook,bies94}.
Applying the chain rule to express things in terms of $\tau$, making use of the identity $\partial_k (\ln \Omega) = \partial_k (\Omega^2)/2 \Omega^2$ yields, after noting the relation $\Omega^2 g_{ij} \left({\Omega^{-2}}{dx^i}/{d\tau}\right)\left({\Omega^{-2}}{dx^j}/{d\tau}\right) = 1$ from the base-space momentum normalisation, the conformal equation of motion
\begin{equation}\label{eq:3D_pure_final}
0 = \Omega^{-4} \left[\frac{d^2 x^i}{d\tau^2} + {}^{(g)}\Gamma^i_{jk} \frac{dx^j}{d\tau} \frac{dx^k}{d\tau} - \frac{1}{2} g^{ik} \partial_k (\Omega^2) \right].
\end{equation}
The bracketed term must vanish if $\Omega$ is itself nowhere vanishing.}

\begin{figure*}
\centering
  \includegraphics[width=\textwidth]{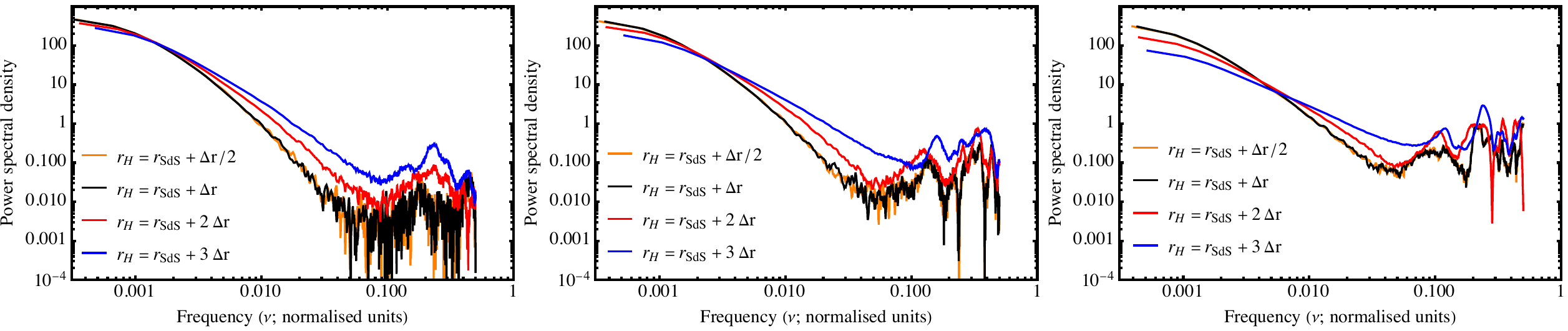}
\caption{Particle fluxes through the inner boundary in different realisations of $N = 3 \times 10^{7}$~walks for $\Delta r = 0.4M$ (left), $\Delta r = 0.5M$ (middle), and $\Delta r = 0.6M$ (right), and different numerical positioning of the event horizon (see plot legends). The global parameters are fixed as in the main text ($\Lambda = 10^{-4}, M = 1, \sigma = 1$). 
  }
  \label{fig:conv1}
\end{figure*}

{Consider now instead a spacetime framework.
For a static spacetime, the 4D line element can be decomposed into a spatial metric, $h_{ij}$, and a lapse function, $N$, through
\begin{equation}
ds^2 = -N^2 dt^2 + h_{ij} dx^i dx^j.
\end{equation}
The time component of the geodesic equations can be integrated once to yield the well-known result
\begin{equation}\label{eq:time_integral}
N^2 \frac{dt}{d\tau} = E,
\end{equation}
for some constant $E$, interpreted as the conserved particle energy.}
{For the spatial components, we can expand out the Christoffel symbols to find the simple relations
\begin{equation}
0 = \frac{d^2 x^i}{d\tau^2} + {}^{(h)}\Gamma^i_{jk} \frac{dx^j}{d\tau} \frac{dx^k}{d\tau} + \frac{1}{2} h^{ik} \partial_k (N^2) \left(\frac{dt}{d\tau}\right)^2,
\end{equation}
which, upon substituting the integration constant from equation \eqref{eq:time_integral}, reduce to
\begin{equation} \label{eq:4dspatial}
0 = \frac{d^2 x^i}{d\tau^2} + {}^{(h)}\Gamma^i_{jk} \frac{dx^j}{d\tau} \frac{dx^k}{d\tau} - \frac{1}{2} h^{ik} \partial_k \left( \frac{E^2}{N^2} \right).
\end{equation}
Comparing equations \eqref{eq:4dspatial} and \eqref{eq:3D_pure_final}, we see that they are equivalent provided that we identify
\begin{equation} \label{eq:omegafn}
\Omega^2 = \frac{E^2}{N^2} - C,
\end{equation}
where $C$ is another constant of integration and we pick $h_{ij} = g_{ij}$. 
Therefore, we can -- at least at a formal level -- recover 4D geodesics starting from the diffusionless limit of a 3D walk via \eqref{eq:somep} provided we use $\Omega^2 h_{ij}$ with \eqref{eq:omegafn} as the base metric rather than $h_{ij}$ directly.}

{As it would clearly be more desirable to recover 4D geodesics in a diffusionless limit, it is tempting to try and reformulate random walks using the conformal metric \eqref{eq:confmap}.
Unfortunately, this leads to a conceptual problem.
Indeed, the conformal scaling $\Omega^2 = {E^2}/{N^2} - C$ enforces a strictly-conserved total energy for the walkers.
This makes sense for a deterministic geodesic as the geometry creates a barrier at the singular point where $N^2 = E^2/C$ (separating, e.g., bound from unbound orbits).
In considering a diffusion process for the Jacobi metric, however, such an $\Omega$ effectively demands that any given particle has a maximum radial extent that is bounded above by the energy defined by its initial state.
This is problematic as it would prevent walkers ever reaching the outer boundary by construction.
It is for this reason we have not attempted to formulate walks in the main text via Jacobi metrics \eqref{eq:confmap}.
Whether such limitations could be overcome by also imposing diffusion equations for $E$ and/or $C$, for example, will be explored in future work.}

\section{Convergence tests} \label{sec:convergence}

\begin{figure}
\centering
  \includegraphics[width=0.485\textwidth]{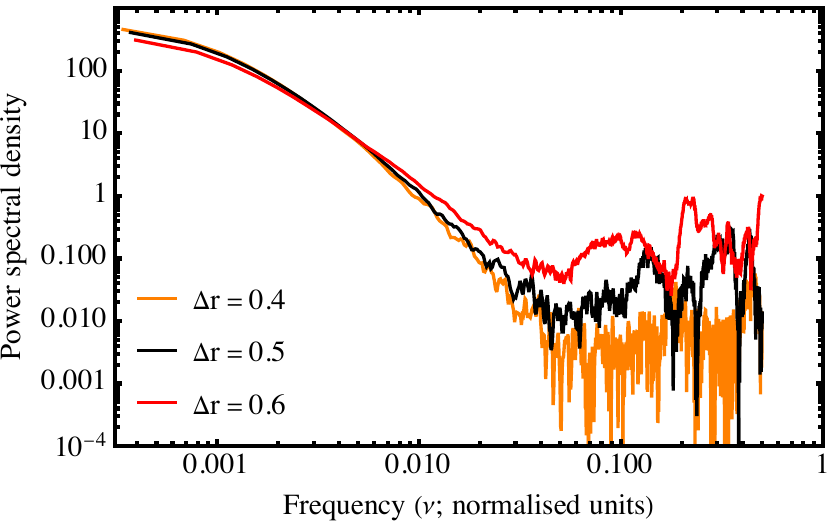}
\caption{Similar to Fig.~\ref{fig:conv1} but instead for a fixed inner boundary $r_{H} = r_{H}^{\rm SDS} + \Delta r$ (i.e., $\epsilon = 1$) and varying spatial step sizes (noting $M=1$) as per the figure legends.
  }
  \label{fig:conv2}
\end{figure}

{We carry out several convergence tests in this Appendix to ensure the reliability of the inner-horizon particle fluxes reported in Section~\ref{sec:accretion}.}

{Since the metric factor, $f(r)$, diverges at the horizon, it is necessary to place an artificial boundary at some slightly-larger radius, $r_{\rm out} > r_{H}$, to absorb walkers. 
We opt to parametrise the edge as $r_{\rm out} = r_{H} + \epsilon \Delta r$ for some $\epsilon >0$, as $\Delta r$ controls the step length and thus the discrete radii that walkers may reach.
In general, by increasing $\epsilon$ we expect a high-pass filtering effect in that large time dilations -- and hence power at low-frequencies -- are removed from the synthetic accretion stream as walkers are absorbed prematurely.
For small $\epsilon$, we expect a wide tail of arrival times because of the scaling relations \eqref{eq:geoweight} and \eqref{eq:dtau}.}

{Such a trend can be seen in Figure~\ref{fig:conv1}, showing first-passage fluxes through the inner boundary for various $\epsilon$.
We see a monotonic decrease in the lowest Fourier frequencies as a function of $\epsilon$; for $\epsilon = 1/2$ and $\Delta r = 0.5 M$ (orange curve, middle panel) the minimum frequency is a factor $\sim 4$ lower than that obtained for $\epsilon = 3$ and the same $\Delta r$. 
Going from the left-to-right panels we increase the numerical step $\Delta r$ to demonstrate that (i) the effect is general for any resolution, and (ii) lower $\Delta r$ also leads to a wider tail as more time dilation can be accrued.
While the overall shape of the PSD is broadly consistent across each of the cases, when $\epsilon > 1$ we see the curves become artificially shallow at low frequencies. 
This is a consequence of the diminishing relativistic effects as we essentially converge to the flat-space case studied in Sec.~\ref{sec:firstpassage} since the region of strongest gravity is excluded \emph{a priori}.
The models with $\epsilon = 1/2$ (orange) and $\epsilon = 1$ (black) are largely indistinguishable for any $\Delta r$.
This fact, combined with the observation that the $\Delta r = 0.4 M$ (left) and $\Delta r = 0.5 M$ (middle) cases are quantitatively similar at low frequencies, indicates convergence for the cases studied in this paper.
} 

{As another test we note that, near the horizon, we can Taylor expand to get $f(r) \approx \kappa(r - r_H) + \mathcal{O}[(r - r_H)^2]$ where $\kappa = f'(r_H)$ is the BH surface gravity. 
By relations \eqref{eq:geoweight1} and \eqref{eq:dtau}, the expected coordinate time to reach the boundary should scale as $\propto 1/\sqrt{\Delta r \epsilon}$.
Such a scaling ought to manifest as a shift in the lowest frequency observed in the large-$N$ PSD.
Figure~\ref{fig:conv2} compares spectra for varying step size but fixed $\epsilon$. 
We see that, in the normalised units adopted here, we have $f_{\rm min} \approx 4 \times 10^{-4}$ (red) and $f_{\rm min} \approx 3 \times 10^{-4}$ (orange) for $\Delta r = 0.4$. 
We thus have the ratio $f_{\rm min}(\Delta r = 0.4)/f_{\rm min}(\Delta r = 0.6) \approx 0.75$ in acceptable agreement with the theoretical value $\sqrt{2/3} \approx 0.82$. 
Comparing the red and black ($\Delta r = 0.5 M$) curves, we find $f_{\rm min}(\Delta r = 0.5)/f_{\rm min}(\Delta r = 0.6) \approx 0.88$ which is close to the theoretical value of $\sqrt{5/6} \approx 0.91$.
}

\end{document}